\documentclass[sigconf]{acmart}
\usepackage{enumerate}
 \usepackage{enumitem}
 \usepackage{caption}
 \usepackage{algorithm}
\usepackage{braket}
\usepackage[noend]{algpseudocode}
\usepackage{subcaption}

\usepackage{array}
\newcolumntype{P}[1]{>{\centering\arraybackslash}p{#1}}
\algdef{SE}[SUBALG]{Indent}{EndIndent}{}{\algorithmicend\ }%
\algtext*{Indent}
\algtext*{EndIndent}

\AtBeginDocument{%
  \providecommand\BibTeX{{%
    \normalfont B\kern-0.5em{\scshape i\kern-0.25em b}\kern-0.8em\TeX}}}

\setcopyright{acmcopyright}
\copyrightyear{2018}
\acmYear{2018}
\acmDOI{10.1145/1122445.1122456}

\acmConference[Woodstock '18]{Woodstock '18: ACM Symposium on Neural
  Gaze Detection}{June 03--05, 2018}{Woodstock, NY}
\acmBooktitle{Woodstock '18: ACM Symposium on Neural Gaze Detection,
  June 03--05, 2018, Woodstock, NY}
\acmPrice{15.00}
\acmISBN{978-1-4503-XXXX-X/18/06}

\begin{document}

\title{Adaptive Epidemic Forecasting and Community Risk Evaluation of COVID-19}



\author{Vishrawas Gopalakrishnan, Sayali Navalekar, Pan Ding, Ryan Hooley, Jacob Miller, Raman Srinivasan, Ajay Deshpande, Xuan Liu, Simone Bianco, James H.\ Kaufman} \authornote{The first two authors have contributed equally}
\affiliation{
  \institution{IBM, USA}
  \country{}
}
\email{
{vishrawas.gopalakrishnan1,sayali.pethe,pan.ding,ryan.hooley}@ibm.com,
{mjacob,rsrin,ajayd,xuanliu,sbianco,jhkauf}@us.ibm.com}

\renewcommand{\shortauthors}{Vishrawas and Sayali, et al.}

\begin{abstract}
Pandemic control measures like lock-down, restrictions on restaurants and gatherings, social-distancing have shown to be effective in curtailing the spread of COVID-19. However, their sustained enforcement has negative economic effects. To craft strategies and policies that reduce the hardship on the people and the economy while being effective against the pandemic, authorities need to understand the disease dynamics at the right geo-spatial granularity. Considering factors like the hospitals' ability to handle the fluctuating demands, evaluating various reopening scenarios, and accurate forecasting of cases are vital to decision making. Towards this end, we present a flexible end-to-end solution that seamlessly integrates public health data with tertiary client data to accurately estimate the risk of reopening a community. At its core lies a state-of-the-art prediction model that auto-captures changing trends in transmission and mobility. Benchmarking against various published baselines confirm the superiority of our forecasting algorithm. Combined with the ability to extend to multiple client-specific requirements and perform deductive reasoning through counter-factual analysis, this solution provides actionable insights to multiple client domains ranging from government to educational institutions, hospitals, and commercial establishments.

\end{abstract}



\begingroup
 \def\UrlFont{\normalsize} 
\maketitle
\endgroup

\section{Introduction}
\label{sec:intro}

Due to the COVID-19 pandemic, it is estimated that half of the world's 3.3 billion global workforces were/are at risk of losing their livelihoods~\cite{WHO_IMPACT} and in the U.S itself, there has been a -11\% change in the number of low-wage category jobs, and overall a -6\% decrease across categories. Similar disruptions are reported across different work sectors with AHA estimating losses of \$202.6 billion for America's hospitals and health systems for the period of March-1 2020 to June-30 2020~\cite{meredith2020preserving} and over 1,300 colleges and universities across all 50 states canceled in-person classes affecting the quality of education as well as leading to financial losses~\cite{ncsl_andrew}. To alleviate these issues, studies have recommended that governments and industries should, as soon as possible, try to resume operations even in a limited capacity and then adapt their operations based on the evolving situation~\cite{shen2020impact}.

 \begin{figure*}[!ht]
 \centering
\includegraphics[height = 5.75cm, width=19cm]{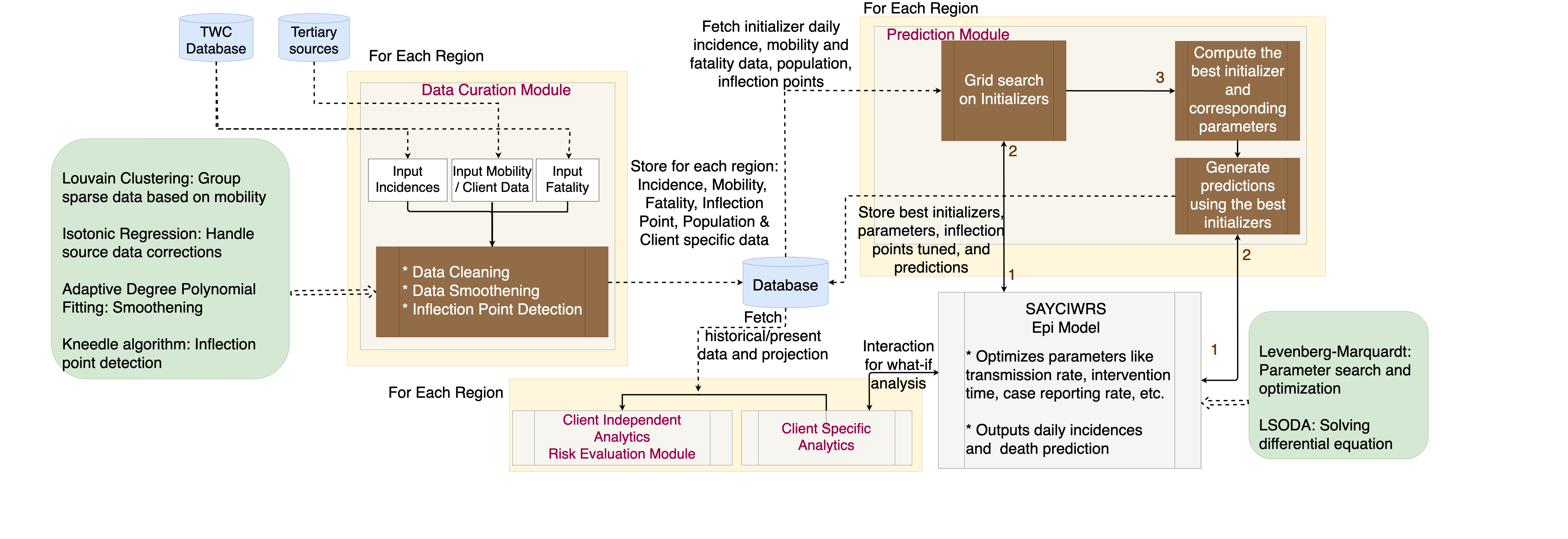}
 \caption{Overview Diagram of the Solution}
 \label{fig:overview}
 \end{figure*} 
   Towards this goal, our solution - part of IBM's Watson Works, provides timely information on the pandemic's current and future state in a region. As a solution framework, with an objective to cater to different specific client domains, it provides actionable insights that support planning safe reopening of offices, crafting school schedules, meeting hospitalization/ICU demands, planning for clinical vaccine trials, etc. Additionally, it has also helped local officials and state government test hypotheses and crafting their intervention strategies for the COVID-19 pandemic. Several efforts like~\cite{Li2020,Oliver2020,brookscomparing,zou2020epidemic,covid2020forecasting} provide forecasts of COVID-19. However, the superiority of our method lies in its ability to provide a solution to:

\begin{itemize}[noitemsep,nolistsep,topsep=2pt]
    \item Model at a hyper-local level.
    \item Translate the current and predicted case load numbers to community reopening risk metrics in accordance with national/regional laws.
    \item Easily adapt to business specific tasks and objectives like hospital/ICU demand projections or identifying appropriate location for vaccine trials.
    \item  Scale globally through distributed processing.
    \item Interact and simulate \textit{what-if} scenarios.
    \item Incorporate tertiary signals apart from cases and deaths data to better estimate the future trends.
\end{itemize}

As shown in Section~\ref{sec:results}, our system consistently performs better in both short and long-term projections, enabling users to take suitable corrective steps. The ``base" prediction model is easily enhanced to handle user-specific tasks by plugging in custom analytics.  For instance, this solution helped a major hospital in south Florida to accurately estimate hospital bed and ICU demands; thus, allowing them to plan/reschedule elective procedures and surgeries, which otherwise could  have got cancelled. It also helped a pharmaceutical company to identify countries for their vaccine's clinical trials. 

Apart from customizing some of the known methodologies for data orchestration, the deployed solution also makes a significant contribution to core epidemiology by providing a template for models to incorporate important tertiary signals without increasing the parameter search space. Further, it also demonstrates a novel way of allowing parameters (e.g., transmission rate) to assume multiple values over the course of the pandemic to better reflect changes in the pandemic, including interventions, policy enactments, and their time-varying efficacy. We also report learnings and insights regarding variation in the relative importance of different input signals over time, the impact of geographic level (county vs.\ state) on the prediction, etc.  



\section{System and Deployment Overview}
\label{sec:overview}


Figure~\ref{fig:overview} provides a holistic view of the data flow of our solution. The solution begins by first connecting to `The Weather Channel' (TWC) API and consuming daily published COVID-19 cases, deaths, and other pertinent statistics. TWC scrapes data from individual federal/state/county government websites and stores data, including individual cases, test positivity rate, hospitalization data, etc. After performing audits on data quality, TWC pushes data into their database for downstream consumption.

The next step in the pipeline is \textit{Data Curation} wherein raw COVID-19 cases and death numbers are pre-processed and merged with tertiary data, including mobility and other client-specific data such as individual hospital and ICU demand data. The latter is optional as the \textit{Client Specific Analysis} module can be disabled. As shown in Figure~\ref{fig:overview}, pre-processing steps include adjusting spatial granularity, handling reporting irregularities, data smoothening, and \emph{inflection point detection}. 
 
The pre-processed data is then passed to the \textit{Prediction Module} which provides case and death projections with upper and lower interval bounds. The current case and death data, along with the projections from the model, are provided to the \textit{Client Independent Analytics} module to calculate various epidemiological metrics and derive a score between 1-6 (lower is better) reflecting the community transmission risk in a region. At the same time, the model prediction output along with supplemental client data drives \textit{Client Specific Analytics} insights like hospitalization/ICU demand or simulating the effect of an event (holiday travel, sports event) or intervention (government restrictions) on the pandemic's transmission rate.

The entire process is wrapped in an Apache NiFi data flow pipeline that invokes relevant modules and manages the data flow. The pipeline is triggered once a day after the latest COVID-19 data is populated by TWC. The `Prediction Module' is triggered once every three days as the projections rarely deviate a lot in such a short period. However, the `Client Independent Analytics', which includes computing community risk, is triggered every day to provide up-to-date statistics like doubling period, 14-day case trends, 1-week, and 2-week ahead community risk levels, etc. The `Client Specific Analytics' is run as per client requirements.

The following sections focus on a pipeline configured for U.S.A data, although the solution itself is applicable globally with relevant changes to account for the availability of tertiary data.

\section{Methodology}
\label{sec:methodology}
This section details various algorithms and steps used in the different modules shown in Figure~\ref{fig:overview} by first introducing the motivation/source of the problem, the solution intuition, and lastly, by providing specific operational examples. 

\subsection{Data Pre-Processing}
\label{sec:preprocessing}
This step entails two important tasks: (i) Denoising the COVID-19 data and (ii) Identifying approximate timestamps of the inflection points in disease dynamics.

\subsubsection{Denoising Public Health Data:}

 \paragraph{\textbf{Geo-Spatial Noise:}}
  Any pandemic is defined by its transmission rate - the average rate at which individuals pass on the infection. This, in turn, is determined by the population contact or \emph{mixing rate} in the region. Assuming a similar contact rate across regions will lead to an incorrect susceptible population. For instance, counties of New York City-Newark-Jersey City, NY-NJ-PA Metropolitan Statistical Area have a higher mixing rate than rural counties of Alaska or North Dakota, which are sparsely populated and self-contained. Thus, the first step is to normalize regions/counties by aggregating them into a self-contained geo-unit of analysis using a metric that reflects the population mixing characteristics. Additionally, it is often found that in an area where there is a major hospital center, reported cases around that area are skewed as patients from nearby counties are reported at the major hospital. For example, COVID cases that occurred in Bronx County, N.Y. were reported as originating in Westchester County, N.Y.  
  \begin{figure}[!t]
     \centering
     \begin{subfigure}[b]{0.27\textwidth}
         \centering
         \includegraphics[height = 2.75cm,width=\textwidth]{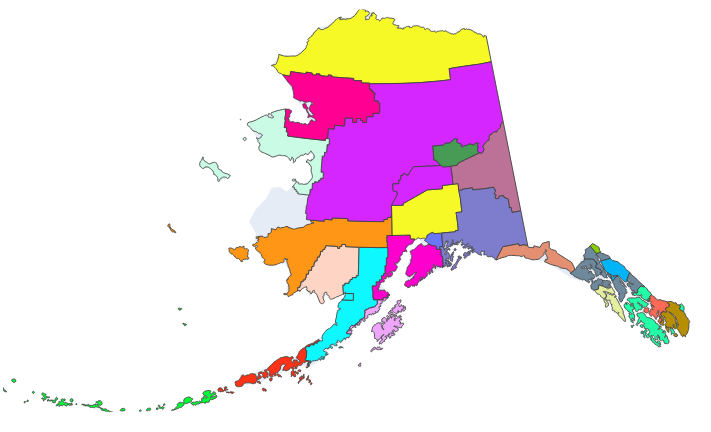}
         \caption{Alaska}
         \label{fig:ak_county}
     \end{subfigure}
     \hfill
     \begin{subfigure}[b]{0.2\textwidth}
         \centering
         \includegraphics[height = 2.75cm, width=\textwidth]{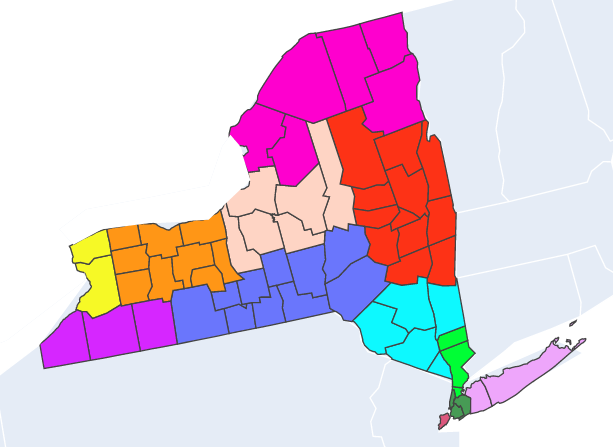}
         \caption{New York}
         \label{fig:NY}
     \end{subfigure}
        \caption{County Grouping by Residence-Work Commute}
        \label{fig:cluster}
\end{figure}
  
  Towards this end, our solution uses Louvain's method of graph clustering~\cite{blondel2008fast} to group adjacent counties together. The algorithm uses the census bureau's residence to workplace commute data 2011-2015\footnote{www.census.gov/data/tables/2015/demo/metro-micro/commuting-flows-2015.html} to create an adjacency matrix where each node is a county and edges represent the strength of commute. Note that we symmetrize the adjacency matrix to reflect the return journey as well. Using this approach, we were able to group 3,132 counties into super-groups of ~685 county clusters. There are about 4.59 counties per cluster, and 19\% of the total clusters are singletons. The intersection of these county clusters with statistical regions like Metropolitan Statistical Areas is high. Further, the clustering method allows us to cover the entire USA as opposed to the limited commercially active zones covered by the census bureau's statistical area maps. Additionally, it provides the flexibility to impose state border conditions on the adjacency matrix; thus, supporting the roll-up of statistics to a state-level when required. Figure~\ref{fig:cluster} shows the county cluster maps of Alaska and New York. 
  
\begin{figure}[!ht]
 \centering
\includegraphics[height = 3.4cm, width = 8.5cm]{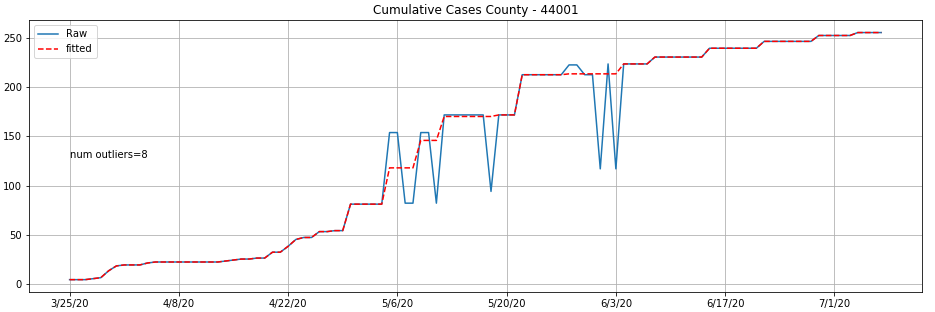}
 \caption{Isotonic Regression on Cumulative Incidence Cases}
 \label{fig:isotonic}
 \end{figure}

  \paragraph{\textbf{Reporting Noise:}}
 Often, public health data sources avoid making retrospective changes to correct reporting errors. As an example, Figure~\ref{fig:isotonic} shows that the \textit{cumulative} incidence data for Bristol County, RI, (FIPS code 44001) is not monotonically increasing. The reason for such aberrations can be numerous, ranging from sources incorrectly assigning new cases to a hospital (instead of the patient's home address) to changes in the way cases are counted. To resolve these issues, we employ Isotonic regression~\cite{chakravarti1989isotonic} defined by:
  \begin{equation}
  \label{equ:isotonic}
      \min \sum_{i=1}^n w_i (x_i - a_i)^2 {\displaystyle {\text{ subject to }}x_{i}\leq x_{j}{\text{ for all }}(i,j)\in E.}
  \end{equation} 
  
  Isotonic regression involves finding a weighted least-squares fit $x \in \mathbb{R}^n$ to a vector $a \in \mathbb {R}^{n}$  with weights vector $w \in \mathbb {R}^{n}$  subject to a set of non-contradictory constraints of the kind $x_{i}\leq x_{j}$. Figure~\ref{fig:isotonic} also shows the fit of Isotonic regression on the noisy data.
  
  While the Isotonic regression ensures the cumulative number of cases is monotonically increasing, it does not safeguard against sudden spikes that arise periodically due to a backlog in lab tests reported (e.g., on the weekend) or correction in previously under-reported numbers. Since we model both daily and cumulative incidences, these fluctuations would affect the model performance. Towards that end, we use an adaptive-degree polynomial filter (ADPF)~\cite{barak1995smoothing} to smoothen the daily cases. It is shown that that ADPF performs nearly as well as the optimally chosen fixed-degree Savitzky-Golay filter and outperforms sub-optimally chosen Savitzky-Golay filters~\cite{barak1995smoothing}. Note that we only calculate \textit{cumulative} deaths as the \textit{daily} death numbers are small, and stochasticity makes daily optimization problematic for deaths. 
  
 \subsubsection{Auto Change Detection in Disease Dynamics: }
 \label{sec:auto-detect-intervention}
 As mentioned earlier, transmission rate defines the dynamics of the pandemic, and this varies over time. The change in the transmission rate (and other disease parameters) primarily stems from Non-Pharmaceutical Interventions (NPIs) and changes in human behavior. NPIs include social distancing practices, stay-at-home orders, night-time movement restrictions, restricted public gatherings, etc. Unlike pharmaceutical interventions like vaccines that reduce the susceptible population by moving individuals from $S$ to a `removed' compartment (e.g., $R$),  NPIs aim to lower the \textit{contact rate} between infectious and susceptible individuals. 
 
 While there are efforts to capture the impact and effectiveness of NPIs in curtailing COVID-19 cases~\cite{suryanarayanan2020wntrac,sharov2020creating,schwartz2020predicting}, there are challenges in using them in a live system. Notably, there are region-specific factors like long weekends or festivities, which affect the disease parameters that are not usually part of the database. There is also an overhead of maintaining an up-to-date list of all interventions. Furthermore, transmission rate changes may happen due to human behavior as well. Lastly, there is often a time delay between the implementation (or relaxation) of an NPI and its effect on the observed cases and deaths. This time delay may be unknown until changes in the cases and deaths are observed. Our decision to auto-detect changes in the transmission rate stems from this last observation. Namely, the entire history of the pandemic in a region is split into multiple timepieces, where each piece represents the observed variable of parameters in that time range. Note that in this work, we focus only on transmission rate changes due to NPIs as there is not yet enough data on vaccination.  

\begin{figure}[!ht]
\centering
\includegraphics[height = 5.75cm, width = 8.5cm]{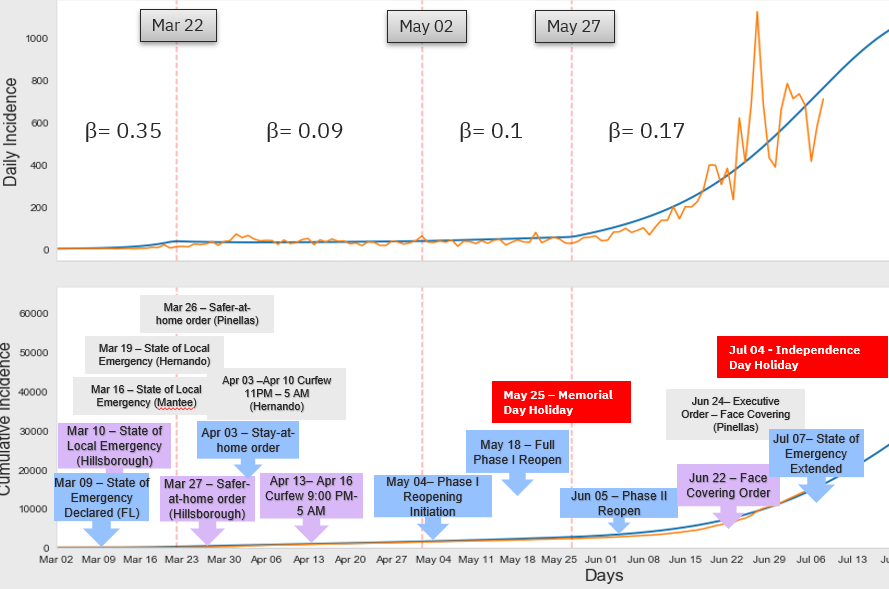}
 \caption{Comparison of auto-detected intervention and fitted transmission rate ($\beta$) with local NPIs and holidays}
 \label{fig:intervention}
\end{figure}
 
 To identify changes in model parameters, we use Kneedle algorithm~\cite{satopaa2011finding} on the COVID-19 case data to detect inflection points. The algorithm is run on smoothened daily case data. If, despite smoothening, the algorithm detects multiple inflection points in close vicinity (within 7 days), they are merged.  We also take into consideration the `significance' of each inflection point, wherein we not only measure the change in the slope of the curve around the inflection point but also ensure that if a `knee' is detected, the subsequent values over two weeks are decreasing (vice-versa for an `elbow'). Figure~\ref{fig:intervention} illustrates the auto-detected intervention for county-level data using our algorithm and superimposes the local government imposed NPI. 
 
 These inflection points provide an approximate \textit{suggestion} to our model (Section~\ref{sec:model_formulation}) regarding times where it might have to (re)adjust parameters to get an optimal fit. Note that this approach is just one of the two parts to dynamically adapt time sensitive parameters like transmission rate. This part handles scenarios where the effect of an intervention (or relaxation) is already visible. The scenario where the effect is anticipated is handled by incorporating mobility as a tertiary source (refer Section~\ref{sec:model_mobility}). 
 
 \subsection{Base Model}
 \label{sec:model}
  
 \subsubsection{Model Basics: }
 \label{sec:model_basics}
We use a compartment structure - a generic framework of models extensively used in epidemiology~\cite{kermack1927contribution,hu2013modeling,balabdaoui2020age,Tsay2020}, to model COVID-19 cases and deaths. 
 A compartment model includes the transition/flow of people across different stages of a disease outbreak. At the minimum, a compartment flow model contains two states - $S$ representing a group of susceptible people in a population and $I$ representing infectious people. Depending on the type of micro-parasitic infection, epidemiologists add compartments based on the organism's life cycle, patient state, and transmission mode. It is worth noting that the ecosystem can span beyond humans; for instance, in the case of Avian Influenza there is an interplay between humans and birds (and mosquitoes) with different sets of parameters. In the case of COVID-19, we would need at minimum two additional compartments to reflect people who recovered ($R$) and died ($D$). It is to be noted that the compartment $R$ should be thought to be as `Removed' compartment; i.e., if the $R$ compartment is initialized with say 20\% of the population in the region, it would then amount to saying that 20\% of the people would never get affected - indirectly accounting for a concept called `herd immunity'. So depending on the biological property of the virus, the `herd immunity' ratio is set - for COVID-19, we initialize $R$ with 20\% of the total population~\cite{fox2020covid}.

\begin{figure}[!t]
 \centering
\includegraphics[height = 3.5cm, width = 8.5cm]{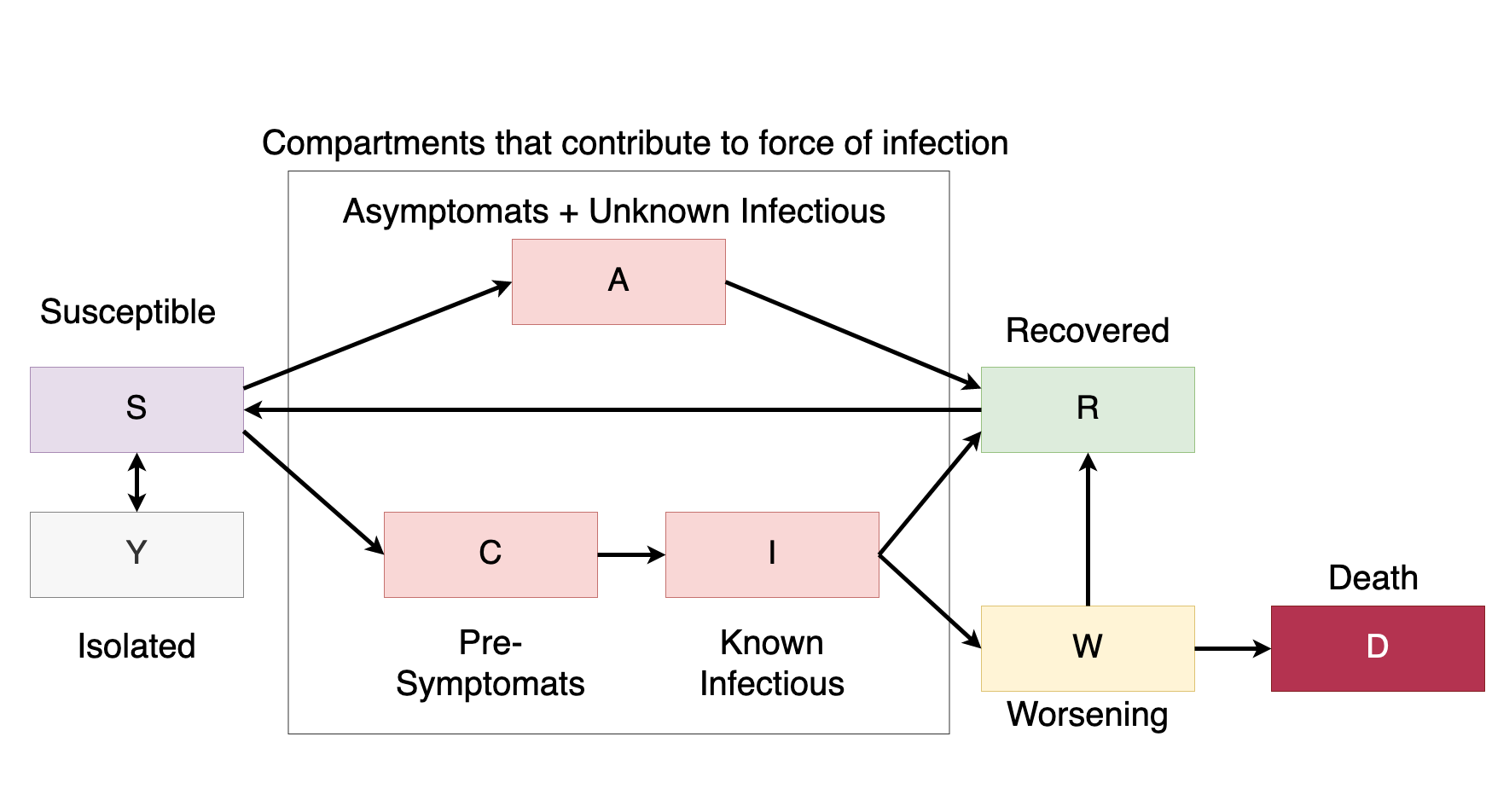}
 \caption{SAYCIWRDS Compartmental Model Flow}
 \label{fig:model}
 \end{figure} 
In our model, we add two additional `infectious' compartments to account for Asymptomats, including unreported infectious people, ($A$) and Pre-symptomats ($C$). Like the normal flu, it is reported~\cite{johansson2021sars} that many people infected by COVID-19 do not exhibit any symptoms. However, they continue to shed viruses and contribute to the force of infection. Conversely, some infected people with mild symptoms may not visit a doctor to be counted in the COVID-19 infected list. The $A$ compartment represents these patients. 

There is an incubation time for an individual to begin exhibiting the symptoms. Even during the time when they are `pre-symptomatic', such people may shed the Sars-CoV-2 virus and contribute to the force of infection. It is necessary to account for this `pre-symptomatic' period in our model. It is important to note that someone being marked COVID-19 positive corresponds to the transition from $C \rightarrow I$. The time it takes for a person to move from $C \rightarrow I$ corresponds to symptom appearance rate ($\alpha$). Normally that represents just the incubation time, but in this case, it also accounts for latency in testing and reporting. 

Figure~\ref{fig:model} represents our compartment model. The $S$ compartment holds the susceptible population after accounting for the percentage of `removed' population used to initialize the $R$ compartment. From the $S$ compartment, an infected person can be either in $A$ or in $C$, which is determined by two things: (a) Likelihood of the person getting infected - represented by the transmission rate ($\beta$), and (b) Likelihood that patient ends up being tested if not showing any symptoms - represented by the case reporting rate ($\xi$). As shown in Figure~\ref{fig:model}, we also introduced a compartment $W$, which does not contribute to the force of infection. The $W$ compartment accounts for patients who become worse and require extra attention and possibly hospitalization. We assume the people in $W$ are isolated in hospital and hence do not contribute to the force of infection. As this base model uses only reported incidence and death data for fitting, $W$ acts as a delay compartment to account for the time delay ($\omega$) between infection and death. People can recover from $A$, $I$, $W$; albeit with different rates $\gamma_A, \gamma_I, \gamma_W$. People enter the death compartment $D$ at a rate of $\mu_D$. COVID-19 being similar to influenza, also has an `immunity loss rate' ($\rho$), denoting the rate at which a person previously infected may lose their immunity and become susceptible again. We set it to 1/(10 months)~\cite{kissler2020projecting}.

\subsubsection{Incorporating Mobility:} 
\label{sec:model_mobility}
 Figure~\ref{fig:model} also shows an isolation compartment ($Y$). The $S$ and $Y$ compartments have a cyclic connection. The idea is to stage $Y$ as a transient state to simulate people who are not mixing with the population. No restriction on population mixing is an assumption made by standard compartment models. However, with the stay-at-home order and decreased mobility/interaction, this assumption is violated. Continuing that assumption would lead to faster depletion of the susceptible population and overestimation of the asymptomats (note that we fit only with respect to cases and deaths, and so the $A$ compartment is indirectly determined by the testing/case reporting rate $\xi$). To account for the population `hidden' from virus exposure and scenarios where effects of relaxation of an intervention are yet to reflect on observed cases (recalling discussions from Section~\ref{sec:auto-detect-intervention}), we set the transition rate from $S \leftrightarrow Y$ to be equal to \textit{rate of change} in the daily mobility. This value is obtained from tertiary sources like Apple\footnote{https://covid19.apple.com/mobility}, Google\footnote{https://www.google.com/covid19/mobility/}, or Descartes Lab\footnote{https://github.com/descarteslabs/DL-COVID-19}. If the normalized difference between two days is positive, it means the mobility is on the rise and the flow from $Y \rightarrow S$ is triggered, else from $S \rightarrow Y$. Thus, at any point in time, the transition between $S$ and $Y$ is always unidirectional. 

 
 The advantage of this model is that we are able to include mobility data without any additional tunable parameter(s). Note that in theory, the isolation compartment $Y$ can be attached to $A$ and $C$ with exits to $R$ and $I$, respectively. However, this will lead to over-complication of the model and an additional parameter $Y_{A} \rightarrow R$. 
 
 \subsubsection{Model Formulation: }
 \label{sec:model_formulation}
Here, we provide the mathematical description of the compartmental flow illustrated in Figure~\ref{fig:model}. These differential equations represent changes in the state of population (each compartment) at any time instance. As the time scale is continuous, we need to add interpolation functions on any of our tertiary sources that directly control the flow - e.g., changes in the mobility that controls $S \leftrightarrow Y$. Let $f_{mob}$ be a function fitted on the mobility, then the equation from $S \leftrightarrow Y$ can be described as:
\begin{eqnarray}
    T_{Y\rightarrow S} = max(0, f_{mob}(t)) \label{equ:return_rate}\\ 
    T_{S\rightarrow Y} = max(0, - 1 * f_{mob}(t)) \label{equ:isolation_rate}
\end{eqnarray}
The fitted function $f_{mob}$ can also be used to extrapolate mobility data to get likely mobility values in the future. Now, the compartmental flow rates can be defined as:
\begin{eqnarray}
\frac{dS}{dt} & = & -\beta * \frac{S * (I + A + C)} {N} + \rho * R + min(Y, T_{Y\rightarrow S} * (S+Y)) \nonumber \\
& &  - min(S,T_{S\rightarrow Y} * (S+Y)) \label{equ:s}\\
\frac{dY}{dt} & = & min(S, T_{S\rightarrow Y} * (S+Y)) - min(Y, T_{Y\rightarrow S} * (S+Y)) \label{equ:y}\\
\frac{dA}{dt} & = & (1 - \xi) * (\beta * S * (I + A + C) / N) - \gamma_{A} * A \label{equ:a}\\
\frac{dC}{dt} & = & (\xi) * (\beta * S * (I + A + C) / N) - \alpha * C \label{equ:c}\\
\frac{dI}{dt} & = & \alpha * C - (\gamma_{I} + \omega) * I \label{equ:i} \\
\frac{dW}{dt} & = & \omega * I - (\mu_{d} + \gamma_{W}) * W \label{equ:w}\\
\frac{dR}{dt} & = & \gamma_{I} * I + \gamma_{A} * A + \gamma_{W} * W - \rho * R \label{equ:r} \\
\frac{dD}{dt} & = & \mu_{D} * W \label{equ:d}
\end{eqnarray}

Note that the $\beta$ in the above equations are time-varying parameters as explained in Section~\ref{sec:auto-detect-intervention} - i.e., there will be one $\beta$ for each of the inflection-detected time intervals, and selecting which beta to modify is based on the time stamp $t$ of the ordinary differential equation (ODE) solver. Recall that the approach in Section~\ref{sec:auto-detect-intervention} only provides an initial estimate for the time of the inflection. Since the ODE solver iterates over time steps, it can find the `optimal' $t$ from the rough estimate to fit the piece-wise curve. Through this, we are able to mimic the natural evolution of the disease, and unlike a regression algorithm, this does not violate any epidemiological principle. We also set $\xi$ ($S \rightarrow A,C$), $\omega$ ($I \rightarrow W$), $\mu_{d}$ ($W \rightarrow D$) to be a time-sensitive parameter similar to $\beta$.

\subsubsection{Model Solver:}
\label{sec:model_solver}
We use an LSODA algorithm~\cite{petzold1983automatic,hindmarsh1983odepack,hindmarsh2005lsoda} to solve the ODEs. LSODA automatically selects between non-stiff (Adams) and stiff (BDF) methods. It uses the non-stiff method initially and dynamically monitors data to decide which method to use. The parameter estimation is done by Levenberg–Marquardt algorithm~\cite{levenberg1944method,marquardt1963algorithm}; selected mainly for its simplicity and speed. The error function to minimize includes daily case NRMSE, cumulative case NRMSE, and cumulative death NRMSE. Because the entire system of equations receives continuous external `shocks' via the mobility values, the solver takes additional time to converge. However, as discussed in Section~\ref{sec:result_mobility}, including this external data provides valuable information and improves performance. 

\subsection{Client Independent Analytics}
\label{sec:client_independent_analysis}
Our solution provides a set of analytics regardless of which client use-case it is serving. These typically include epidemiological metrics like $R_0$, doubling time, 14-day rolling average, etc. Amongst these, $R_0$ is one of great importance. It provides a likelihood of how much an infected individual is likely to infect others. It is directly connected to the transmission rate ($\beta$) and recovery time ($\gamma$). However, it is also model specific. Solving Equations~\ref{equ:a} to Equation~\ref{equ:d}, under steady-state, we derive $R_0$: 
\begin{equation}
    \begin{aligned}[b]
        R_0 & = \frac{\beta }{\gamma_A} \left\{ 1+ \left(\frac{\gamma_A}{\gamma_I+\omega}\right) - \left[1-\frac{\gamma_A}{\alpha }\right]\xi \right\}
    \end{aligned}
\label{equ:r0}
\end{equation}  
where the $\beta$, $\gamma_A$, and $\gamma_I$ in the above equation refer to the first fitted value of $\beta$, $\gamma_A$, and $\gamma_I$.
Recall that we fit multiple $\beta$ mimicking the different transmission rates over the course of the pandemic in a region.  If we calculate $R$ at each of those different times, then using the fraction of susceptibles during each of those times, we get $R-$effective ($R_{eff}$). $R_t$ is used to denote the latest $R_{eff}$.

Together with doubling time (Equation~\ref{equ:doubleTime}), these two statistics provide an accurate sense of the pandemic situation on the ground. 
\begin{equation}
\label{equ:doubleTime}
    {\text{doubling time}}={\frac {\ln(2)}{\text{growth\_rate}}}
\end{equation}
where $growth\_rate$ is defined as 
\begin{equation}
    {\text{growth\_rate}} = \frac {\ln \left(N(t)/N(0)\right)}{t}
\end{equation}

\subsubsection{Community Risk Evaluation}
\label{sec:model_community_risk_eval}
The community risk evaluation uses a rule-based approach to convert the epidemiological metrics into risk scores for a region between 1-6, with 1 being the safest. The features that determine the score include recent incidence trends, the case prevalence in the population, and local government reopening policy. Of these, the case prevalence and government reopening policies determine the manual time-dependant adjustable thresholds in our algorithm. Algorithm~\ref{algo:comm_risk} describes the algorithm to determine the community risk. In our algorithm, there are 3 thresholds - a soft ($\kappa$), a hard ($\lambda$), and a flat threshold ($\tau$), that help merge the three features into one value.  For the U.S.A, based on White House reopening guidelines, we identified $\kappa$ to be 10 cases/100K, $\lambda$ to be 5 cases/100K, and $\tau$ to be 2 cases/100K below which week-over-week change is considered flat. These values were adjusted to account for local reopening guidelines like the traffic light system used in Mexico and 5 alert levels used in the U.K.

\begin{algorithm}[!t]
\caption{\sc Community Risk Determination Algorithm}
\label{algo:comm_risk}
\begin{algorithmic}[1]
\State {\bf Input:} 21 days historical data ($A_1$, $A_2$, $A_3$, with $A_3$ being the latest weekly average) and predictions ($A^{'}_1$,$A^{'}_2$,$A^{'}_3$), $\kappa$, $\lambda$, $\tau$

\State {\bf Output:} Current and 3 weeks projected community risk scores

\State is\_strict\_decr = $A_3 < A_2 \land A_2 < A_1$
\State is\_strict\_incr = $A_3 > A_2 \land A_2 > A_1$

\If{ $A_3$, $A_2$, $A_1$ < $\kappa$}
    \If{$A_3$, $A_2$<$\lambda$} 
            \State is\_flat = $abs(A_3-A_2), abs(A_2-A_1), abs(A_3-A_1)<=\tau$
            \State is\_flat\_decr = $abs(A_2 - A_1) <= \tau \land (A_3 < A_2)$      
            \If{$is\_flat\lor is\_strict\_decr \lor is\_flat\_decr$}
                   \State curr\_risk\_score = 1.0
            \EndIf
            \State \textbf{ else} curr\_risk\_score = 2.0  
    \EndIf
    \State \textbf{else if} $is\_strict\_decr$ \textbf{then} return curr\_risk\_score = 2.0
    \State \textbf{else} curr\_risk\_score = 3.0
\Else
\State \textbf{if} $is\_strict\_decr$ \textbf{then} curr\_risk\_score = 4.0
\State \textbf{else if } $is\_strict\_incr$ \textbf{then} curr\_risk\_score = 6.0
\State \textbf{else} curr\_risk\_score = 5.0
    \EndIf
\State Repeat Steps 3-17 for $A^{'}_1$,$A^{'}_2$,$A^{'}_3$ to get projected risk scores     
\end{algorithmic}
\end{algorithm}
  
\subsection{Client Specific Analytics}
\label{sec:client_specific_analysis}
 
\begin{figure}[!ht]
 \centering
\includegraphics[height = 1.75cm, width = 8cm]{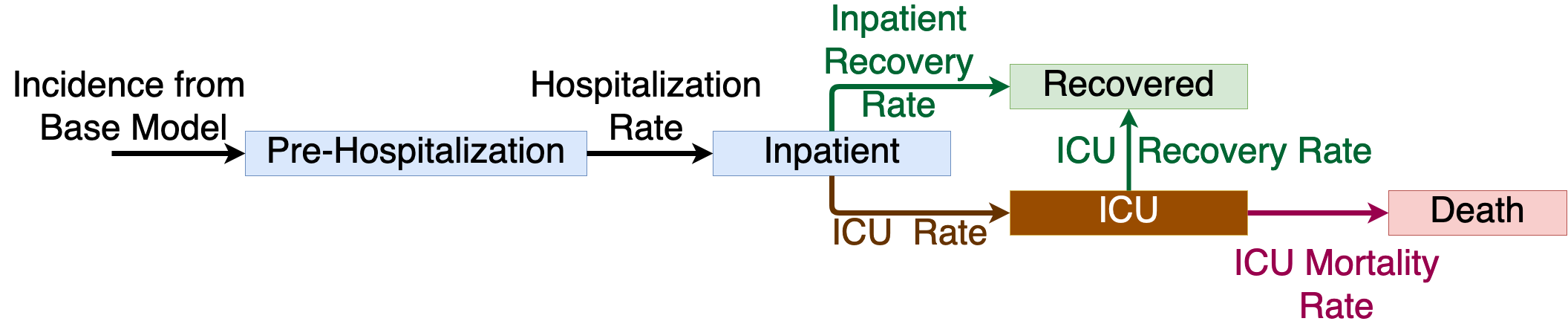}
 \caption{Hospitalization Model Add-on}
 \label{fig:model_hospital}
 \end{figure}

 As mentioned above, our base model could be easily extended for client-specific use cases. In one of the client engagements, we augmented the base model with hospitalization data to predict both hospitalization and ICU demand. Figure~\ref{fig:model_hospital} shows the model extension. The input to this extension is the predicted incidence from our base model. Note that we do not use the predicted deaths from the base model \textit{as-is}, because in this case, we need to infer the mortality rate from the ICU training data. The hospitalization, ICU, and death rates are learnt using the same infrastructure used in solving the base model. The equations can be derived similarly to our base model described in Section~\ref{sec:model_formulation}.

 As shown in Section~\ref{sec:results_counterfactual}, extending to simulate \textit{what-if} scenarios are also straightforward, as one can adjust the learnt parameters and run the model to solve the ODEs in a `scoring' mode.

\begin{table*}[!t]
 \scriptsize
\begin{tabular}{|c|c|c|c|c|c|c|c|c|}
\hline
     & \multicolumn{3}{c|}{AUGUST - 3rd}                                                                                & \multicolumn{3}{c|}{OCTOBER - 26th}                                                                                        \\ \hline
Rank & Week-1          & Week-1+2        & Week-1+2+3                                                    & Week-1               & Week-1+2                                                      & Week-1+2+3            \\ \hline
1    & IBM (0.24)      & IBM (0.23)      & LANL (0.26)                                                     & IBM (0.19)           & IBM (0.162)                                          &     IBM (0.223)          \\ \hline
2    & LANL (0.42)     & LANL (0.27)     & IBM (0.27)                                                     & ESG (0.216)          & Oliver Wyman (0.174)                                                   & Oliver Wyman (0.228)                        \\ \hline
3    & LNQ (0.52)      & LNQ (0.31)      & LNQ (0.31)                                                      & COVID-19 Sim (0.217) & Karlen (0.1851)                                               & LNQ (0.244)                       \\ \hline
4    & DDS-NBDS (0.54) & DDS-NBDS (0.34) & DDS-NBDS (0.34)                                                 & UCLA (0.226)         & LNQ (0.1853)                                                  & Karlen (0.248)              \\ \hline
5    & ERDC (0.56)     & ERDC (0.36)     & USC (0.35)       & UGA-CEID (0.23)      &  Google-HSPH (0.22) & DDS (0.27)                         \\ \hline
\end{tabular}
\caption{Incidence MAPE }
\label{tab:result_mape_inci}

\end{table*}

\section{Results}
\label{sec:results}
Having described the relevant algorithms and modules, we now present a detailed evaluation analysis of the results of the various modules in the system. Our experiments can be broken down into two broad categories wherein we first measure the prediction performance of our base/core system, and then through various user story scenarios, explain the performance, usefulness, and dexterity of the various analytic modules. 

\paragraph{Base Module Evaluation Scheme: }  We use Mean Absolute Percentage Errors (MAPE) to quantitatively measure the  prediction performance. We compare our results with the baselines presented in CDC\footnote{www.cdc.gov/coronavirus/2019-ncov/covid-data/mathematical-modeling.html}. To avoid bias in measuring performance, we measure the accuracy at both the falling phase (end of first wave \textasciitilde Aug-3$^{rd}$) and the rising curve (start of the second wave \textasciitilde Oct-26$^{th}$). Note that CDC did not report case prediction before Aug-1$^{st}$. We also perform a time series ablation study to measure the importance of mobility data and then wrap up the base module analysis by presenting geo-spatial and running time analysis.  

\paragraph{Analytic Module Evaluation Scheme: }
We consider real use-cases to characterize the performance of the `Client Independent Analytics' and `Client Specific Analytics' modules. In one of the client engagements pertaining to the state of Rhode-Island, we highlight how the compartment model helps track the disease progression in the region and compare the validity of the estimated community risk scores with the situation on the ground. Through our support for a major hospital in Tampa, FL, we evaluate the hospital and ICU demand prediction module. As part of our engagement, we also simulate what-if scenarios in response to the Thanksgiving and holiday season, enumerating the effectiveness of various counter-measures. 
\subsection{Base Model Evaluation}
\subsubsection{Prediction Evaluation: }
 Table~\ref{tab:result_mape_inci} shows the cumulative weekly incidence MAPEs calculated from the end of training dates (August-3$^{rd}$ and October-26$^{th}$). As we can see, the `IBM' system consistently comes in top-3, and for the week-1 and week-1+2, it is at the top. The numbers in the bracket indicate the MAPE, and we can see that for the week-1, the difference is substantial. Note that for October, the rest of the top-5 baselines do not make it into top-5 for the remaining weeks. It is also to be noted that some of the methodologies like Oliver Wyman started publishing their results in September and hence do not appear in the August comparison. 
 
 Table~\ref{tab:result_mape_death} shows the cumulative deaths MAPE. In the interest of space, we report only the October-26$^{th}$ value. We observe a similar trend like Table~\ref{tab:result_mape_inci} where our model consistently appears in top-2. 
 
 We also noted that some reported methodologies predict \textit{only} cases or deaths (e.g., YYG). Based on Table~\ref{tab:result_mape_inci} and ~\ref{tab:result_mape_death}, IBM and Oliver Wyman are comparable models with IBM doing particularly well in the 1$^{st}$ week. There are two important points to remember while interpreting these tables.  First, death numbers are substantially smaller than incidence, so the sensitivity of MAPE to incorrect predictions is higher. Second, as the MAPE is calculated based on weekly values, a small deviation in the reported ground truth numbers at the end of an evaluation week may adversely impact the MAPE calculation itself.
 
 \begin{table}[!t]
 \scriptsize
\begin{tabular}{|c|P{2cm}|P{2.5cm}|P{2cm}|P{2cm}|}
\hline
Rank & Week-1                & Week-1+2               & Week-1+2+3           \\ \hline
1    & IBM (0.0146)          & Columbia (0.0146)      & IBM (0.026)      \\ \hline
2    & Oliver Wyman (0.0167) & IBM (0.0189)           &  MSRA (0.36)        \\ \hline
3    & CovidComplete (0.017) & MOBS (0.0245)          & Oliver Wyman (0.037) \\ \hline
4    & MOBS (0.01745)        & CovidComplete (0.0247) & Karlen (0.038)      \\ \hline
5    & USC (0.01779)         & Oliver Wyman (0.025)   & UMass-MB (0.0384)    \\ \hline
\end{tabular}
\caption{Death MAPE for October-26$^{th}$ }
\label{tab:result_mape_death}
\end{table} 
\subsubsection{Impact of Mobility: }
\label{sec:result_mobility}
\begin{figure}[!t]
 \centering
\includegraphics[height = 3.5cm, width = 8.5cm]{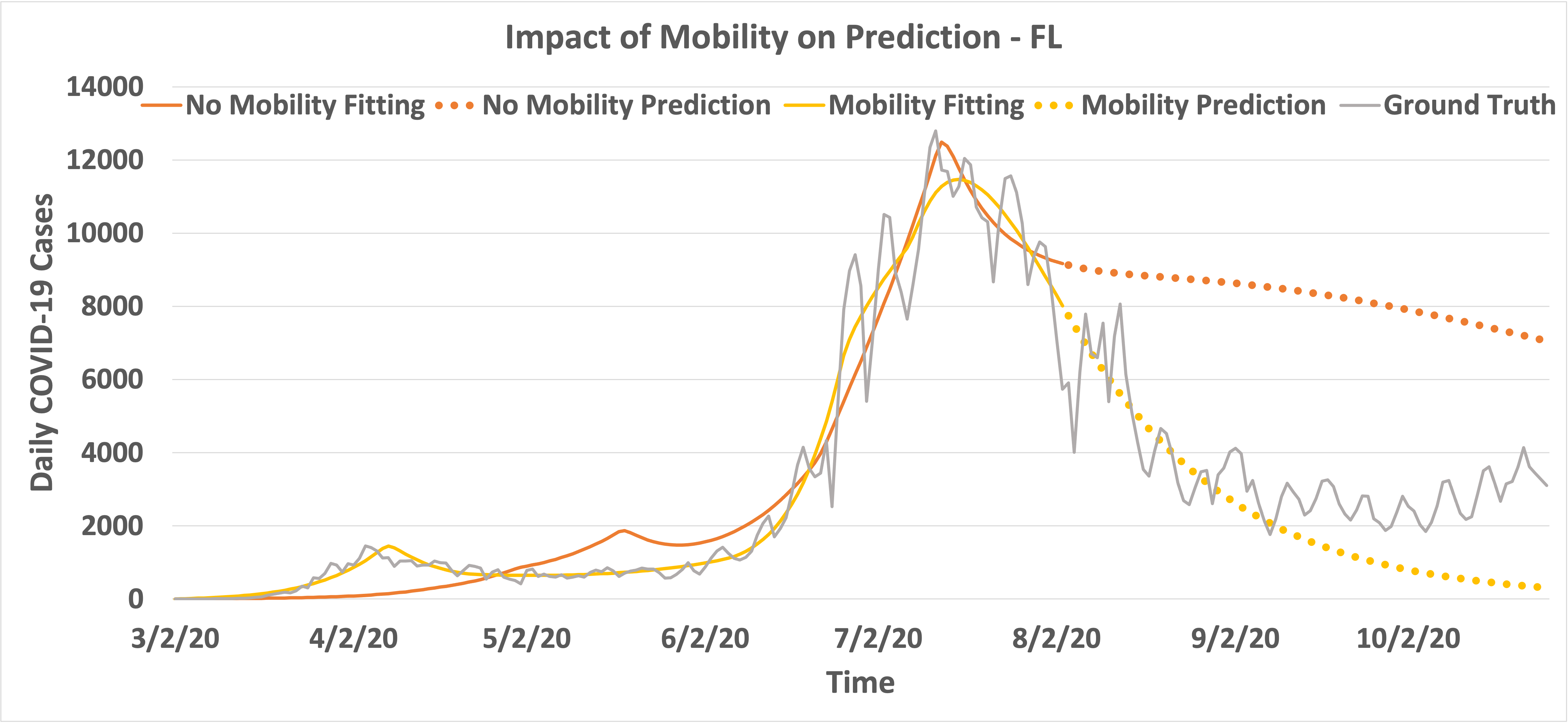}
 \caption{Predictions with and without Mobility}
 \label{fig:mobility_fl}
 \end{figure}

Table~\ref{tab:mobility} compares the MAPE of the model with and without mobility for multiple runs. While there is an improvement across the board, we note a 11\% improvement in the pandemic's early days. Figure~\ref{fig:mobility_fl} shows how the mobility data is able to adjust the trajectory more accurately. While some studies have already suggested that mobility is a leading indicator for COVID-19 cases~\cite{miller2020mobility}, in this work, we further show how to incorporate that information in practice and its usefulness in not only improving the accuracy but also understanding the population mixing and the susceptible pool. We also observed that the contribution of mobility data towards prediction has waned over time.  We suspect this may be due to increased awareness, mask adoption, and increasing herd immunity (i.e., as the population develops antibodies from infection or vaccination). Thus, it is our learning that using mobility data, if available, is a must to get accurate predictions during the early stages of a pandemic. It provides a more precise estimate of susceptibles and the magnitude of the potential second wave.

\begin{table}[!t]
\scriptsize
\begin{tabular}{|c|c|c|c|c|}
\hline
        & Week-1 & Week-2 & Week-3 & Week-4 \\ \hline
Aug-3   & 9.25   & 10.38  & 12.78  & 15.29  \\ \hline
Sept-14 & 2.63   & 1.82   & 1.54   & 1.12   \\ \hline
Oct-26  & 2      & 1.59   & 0.71   & 0.36   \\ \hline
\end{tabular}
\caption{Percentage Improvement In Incidence MAPE by Incorporating Mobility Data}
\label{tab:mobility}
\end{table}

 \subsubsection{Impact of Geo-Location:} The choice of geo-unit for analysis is important. As shown in Table~\ref{tab:result_geo}, modeling at a state level as opposed to aggregating the projections from county-cluster leads to sub-par MAPE. This is because the \textit{local} intricacies of population mixing are lost when the entire state is considered a well-mixed unit.  Thus, crafting an accurate policy at a higher geo-level requires hyper-local modeling. A similar conclusion also holds true when using the state projections to estimate county clusters using area population to drill down. In general, we recommend modeling at county-clusters granularity during the pandemic's initial stages to avoid data sparsity issues. Once the counties within the cluster enter the exponential growth phase (vs.\ sporadic or intermittent case reports), modeling at the county level will prove beneficial, albeit with a trade-off on the running time.
\begin{table}[!t]
\scriptsize
\begin{tabular}{|c|c|c|c|}
\hline
\multicolumn{2}{|P{4cm}|}{Average[(State MAPE)/ (County Cluster Aggregated to State MAPE)]} & \multicolumn{2}{P{4cm}|}{Average[(State Drilled Down to County Cluster MAPE)/(County  Cluster MAPE)]} \\ \hline
Incidence                                   & Deaths                                  & Incidence                                        & Death                                       \\ \hline
1.44                                        & 1.105                                   & 1.818                                            & 9.27                                        \\ \hline
\end{tabular}
\caption{Comparison of Effect of Geo-Unit on Incidence and Death MAPEs}
\label{tab:result_geo}
\end{table}

\subsubsection{Running Time Efficiency:}
 The running time of the pipeline is determined by the choice of parameter optimization algorithm and the external ``shocks" introduced through tertiary sources like mobility data. These shocks cause a sudden change in the equilibrium of the system. Furthermore, with the increase in the number of timestamps (and consequently the number of time-sensitive parameters), the effect is non-linear. Table~\ref{tab:results_time_mob} shows the running time of finding the best parameters for 20 initializer combinations. 
  \begin{table}[!t]
\scriptsize
\begin{tabular}{|c|c|c|c|c|}
\hline
Time (mins) & 05/01 & 08/03  & 09/14  & 10/26  \\ \hline
No Mobility & 1.317 & 13.33  & 19.82  & 30.03  \\ \hline
Mobility    & 40.61 & 171.54 & 273.03 & 404.33 \\ \hline
\end{tabular}
\caption{Running Time Comparison With and Without Mobility Data for 20 Initializer Combinations}
\label{tab:results_time_mob}
\end{table}

 The time-sensitive parameters also make our ODEs stiff at certain places and our experiments with different differential equation solvers, including Runge-Kutta, BDF, and LSODA, showed that LSODA gave the best performance.
 
 We also evaluated different parameter estimation algorithms. Table~\ref{tab:results_time_parameter} shows the efficiency of the different algorithms with respect to Levenberg–Marquardt algorithm, determined to be the best choice. Note that we ensured convergence of approximately the same training error when measuring the running time efficiency.

 \begin{table}[!t]
 \scriptsize
\begin{tabular}{|c|c|c|c|c|}
\hline
                    & Levenberg – Marquardt & SLSQP & Nelder-Mead & BFGS  \\ \hline
Times Slower & 1X                  & 3.5X          & 5X   & 5X  \\ \hline
\end{tabular}
\caption{Running Time Comparison of Different Parameter Estimation Algorithms}
\label{tab:results_time_parameter}
\end{table}

\subsection{Analytics Module Evaluation}
\subsubsection{Community Risk Prediction - Case Study:}
\begin{figure}[!t]
         \centering
         \includegraphics[height=4cm,width=8.5cm]{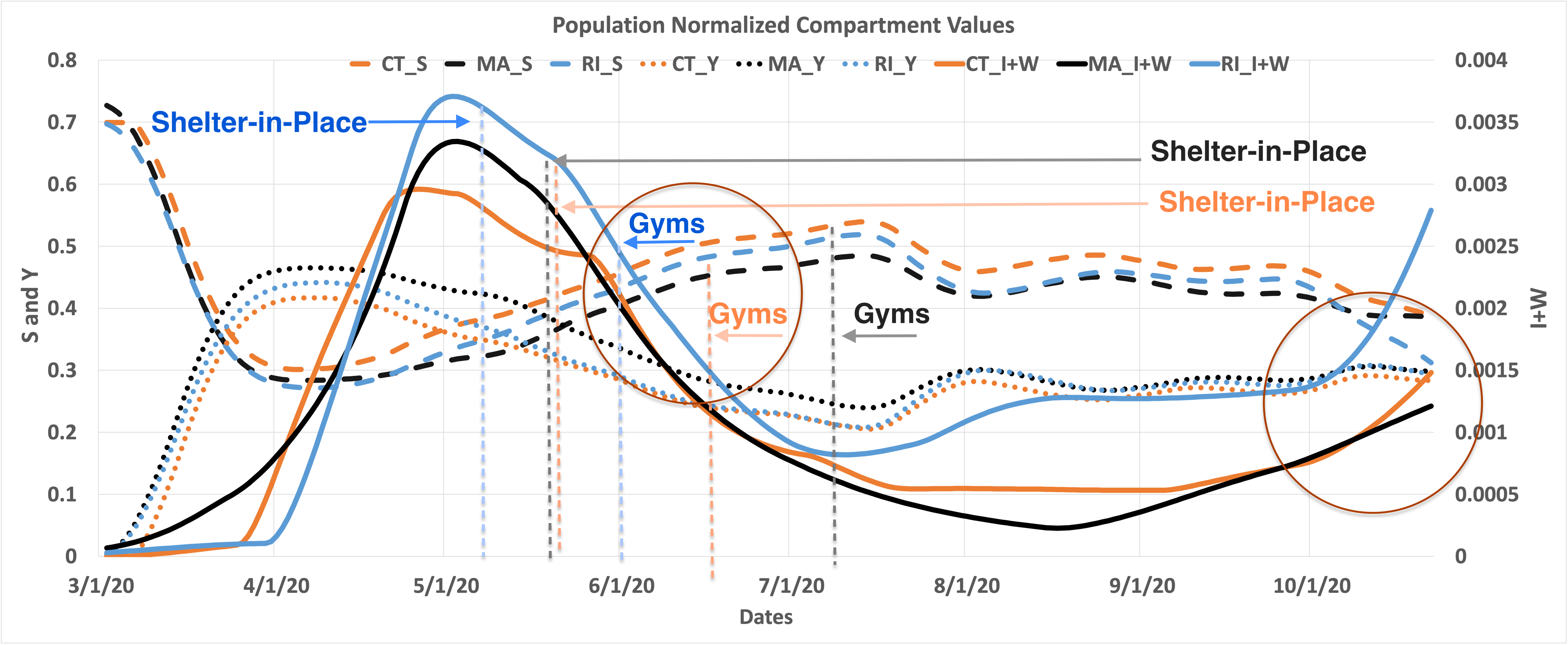}
        \caption{Population Normalized $S$, $I$ and $I+W$ Compartment Values for RI, CT \& MA with reopening dates}
        \label{fig:result_ri}
\end{figure}

 The tri-state region of Rhode Island, Connecticut, and Massachusetts share a common border and worked in tandem to enact restrictions in March-April to control the pandemic. However, as shown in Figure~\ref{fig:result_ri}, their end date of restrictions are not similar. In the majority of the cases, RI had relaxed their restriction 2-3 weeks earlier.
 
 Based on the caseloads and predictions at the time, our solution estimated that in mid-June early-July, CT and MA were safe to reopen with risk values of around 1-2. RI was still marked as unsafe with a score of 5. These scores were validated when the cases started increasing for RI in August while CT and MA were still doing okay. 
 
 To validate these numbers, we examine the model's compartmental values around the time of June reopening. 
  Figure~\ref{fig:result_ri} shows the population normalized \textit{fitted}  compartment values of $S$ (susceptible), $Y$ (isolation), and $I+W$ (total infectious) from March-1$^{st}$ to October-20$^{th}$. It also superimposes the end date of the restrictions. Comparing the prevalence (normalized infection load) across states at the time of reopening, we can observe that RI's prevalence was high, and it stayed relatively high throughout the summer. Positivity rates also remained high (note that this also depends on state policy on who gets tested). Looking at the normalized fraction in $S$ and $Y$, we see that all states are comparable. Thus, the higher numbers in RI (from August onwards) have to do with the high sustained prevalence, and this was in part because of early relaxation of restrictions (15 days to 1 month in advance). This not only validates that the system correctly estimated the risk scores for RI but also shows that CT would most likely have a second wave earlier than MA again. As expected, the risk score of RI further worsened to 6, while CT and MA eventually slipped to risk level 6 from 3 on October-27$^{th}$ and October-29$^{th}$, respectively. 
  
\subsubsection{Predicting Hospitalization and ICU Demand:}
\label{sec:results_hospital}
\begin{figure}[!t]
 \centering
\includegraphics[height = 4cm, width = 8.5cm]{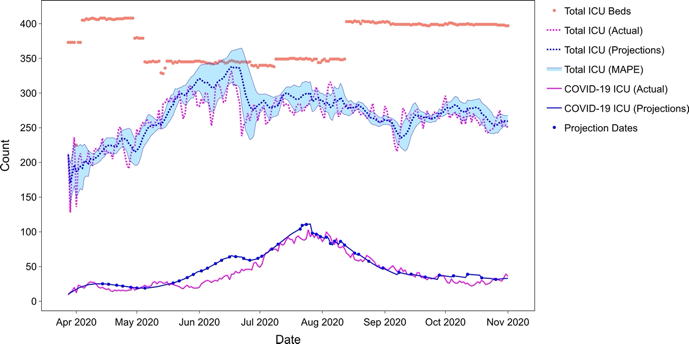}
 \caption{Predicting Hospitalization \& ICU Demands }
 \label{fig:result_hospital}
 \end{figure}

Figure~\ref{fig:result_hospital} shows the fitting and the prediction of hospitalization and ICU beds for Hillsborough County, FL. The figure is for an engagement between April to November 2020, where we provided projections at a regular cadence (indicated in the legend). The model accuracy predicted a peak incidence on 7/20/2020 and peak ICU on 7/26/2020. The client used these projections to chart a hospital capacity forecast and take steps to handle the surge. The blue zone above and below the total ICU prediction represents the MAPE (5.9\% on average). Based on the projections, the hospital decided not to cancel other elective procedures and surgeries in the time period shown. 

 \subsubsection{Counterfactual Analysis - Hillsborough Use Case: }
 \label{sec:results_counterfactual}
\begin{figure}[!t]
 \centering
\includegraphics[height = 3.7cm, width = 8.5cm]{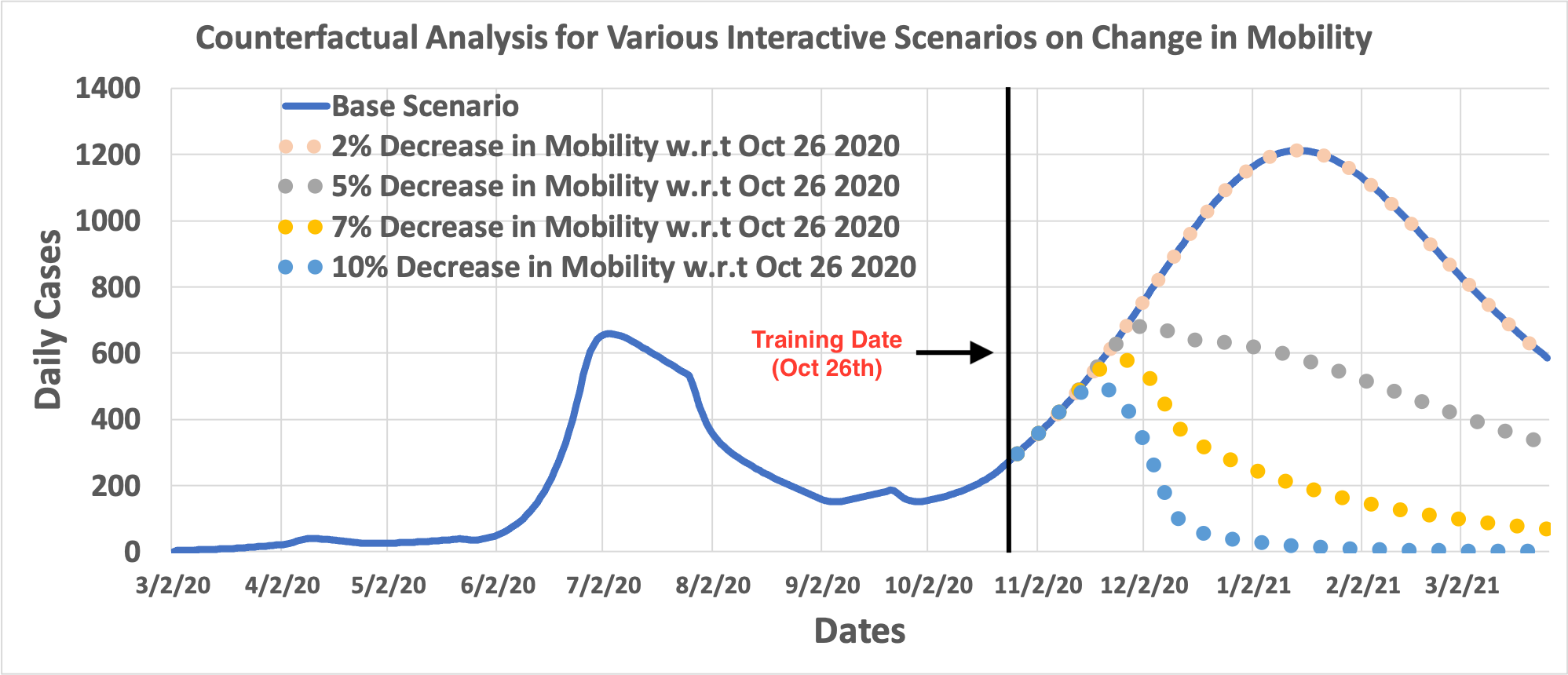}
 \caption{Simulating \textit{what-if} Scenarios with Different Mobility Restrictions - Hillsborough County, FL}
 \label{fig:result_counterfactual}
 \end{figure}

 Counterfactual analysis is important in simulating \textit{what-if} scenarios. Clients, especially local authorities like hospitals or government, would like to understand the effects of future events (including seasonal holidays) on the state of the pandemic. At the same time, it also allows them to plan and take remedial steps. For instance,  Figure~\ref{fig:result_counterfactual} shows the base scenario wherein we show the prediction based on mobility and training as of October-$26^{th}$, 2020 for Hillsborough County, FL. With the Thanksgiving holidays around the end of November, local government officials may be interested in knowing ways to reduce the spread of COVID-19. As mentioned earlier, the transmission of a virus is primarily based on population mixing, and one potent way to curb the population mixing is to enforce lock-downs. However, lock-downs have a detrimental effect on the economy, and so one needs to thoroughly understand the implication of different degrees of enforcement on the spread so that one can take the most appropriate step without overreacting.
 
 Figure~\ref{fig:result_counterfactual} shows the change in daily COVID-19 cases with varying mobility. Note that we do not change any other parameters obtained from the base scenario. The most potent intervention for this state of the epidemic appears to be a 10\% reduction in the mobility values observed on October-$26^{th}$. However, that would require a significant level of curtailment. By comparing the historical case trends, the 7\% decrement scenario matches structurally to the events observed in July. A 5\% reduction achieves the effect of stay-at-home; however, it leads to a slower decline in the subsequent epidemic wave (gray curve). The figure also reveals a minimum threshold level for mobility curtailment of  >\textasciitilde 2\%.  In the case of 2\% reduction, although the mobility value has decreased, there is still an insufficient movement of people into the $Y$ compartment to significantly reduce future incidence over the base scenario. Only for higher values does mobility reduction impact future cases.
 This form of \textit{what-if} analysis allows the authorities to assess multiple possible interventions and provides insight into their relative efficacies based on measurements of the effects of past NPI's. 
 
\section{Deployment and Conclusion}
\begin{figure}[!t]
 \centering
\includegraphics[height = 5cm, width = 8.5cm]{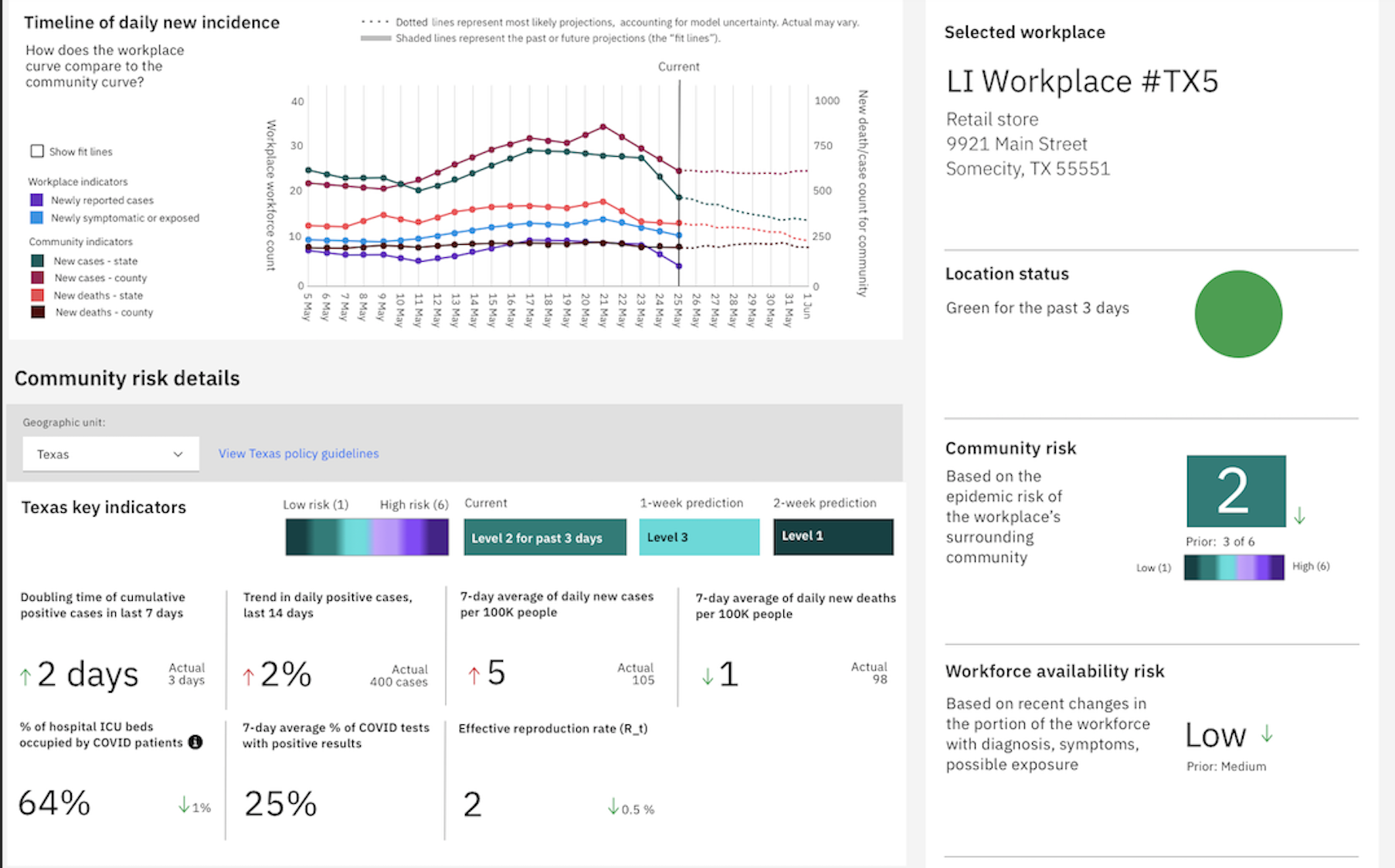}
 \caption{User Interface of the Solution}
 \label{fig:ui_snapshot}
 \end{figure}

The solution uses Python for modeling and JAVA for data orchestration. We built a Docker container and deployed it in IBM Cloud Kubernetes Cluster so that each geo-unit can be processed by an image independently. Apache NiFi orchestrates the entire data flow. There are two NiFi pipelines, one for the core epidemiological module, which is run once in three days. The second is for the `Client Independent Analytics', a.k.a community risk pipeline, which is run nightly.  A manifest configuration file containing the list of geo-ids and any list of hyper-parameters like model compartment initializer values is used to spin one new pod on the cluster for the data flow of a geo-unit. This step is throttled to 18 geo-units at a time due to cluster performance limitations. Note that each pod itself is highly parallelized to solve multiple compartment initializer combinations concurrently. On completion of a pod, the NiFi pipeline stores the output of the epidemiology run in an IBM COS bucket. The Community Risk NiFi pipeline, which is independently triggered nightly, collects the results from the COS bucket and then triggers the process to download the latest case data and merge it with the epidemiological predictions. After computing the statistics and risk scores, it finally pushes the output to DB2 On Cloud through a Data Access Layer (DAL) for clients to consume. The `Client Specific Analytics' has a separate private pipeline that uses these outputs with any additional client data. Figure~\ref{fig:ui_snapshot} shows the final UI presentation to the end-user. 

 In the near future, we intend to extend this solution with additional signals like vaccines, demography, regional comorbidity factors, etc. We strongly believe that the solution and findings of this paper would render epidemiologists and decision-makers better equipped under similar conditions in the future.  
 

\bibliographystyle{ACM-Reference-Format}
\bibliography{reference}

\end{document}